\documentclass[osajnl,twocolumn,showpacs,superscriptaddress,10pt]{revtex4-1}
\usepackage{amsmath,amssymb,graphicx}
\usepackage{color}

\begin{document}

\title{Pulse propagation in one-dimensional disordered photonic crystals: Interplay of disorder with instantaneous and relaxing nonlinearities}

\author{Denis V. Novitsky}
\email{Corresponding author: dvnovitsky@tut.by} \affiliation{B. I.
Stepanov Institute of Physics, National Academy of Sciences of
Belarus, Nezavisimosti Avenue 68, BY-220072 Minsk, Belarus}

\begin{abstract}
Propagation of ultrashort light pulses in disordered multilayers is
studied by using numerical simulations in time domain. We consider
cases of instantaneous and noninstantaneous Kerr nonlinearities of
the structure materials. The competitive nature of disorder and
nonlinearity is revealed on the long and short timescales. We also
pay special attention to the effect of pulse self-trapping in the
photonic crystal with relaxing nonlinearity and show the dependence
of this effect on the level of disorder. We believe that the results
reported here will be useful not only in the field of optics but
also from the standpoint of the general problem of classical waves
propagation in nonlinear disordered periodic media.
\end{abstract}


\maketitle

\section{Introduction}

Disordered photonic structures have become the object of active
study in recent years. Such interest is substantially due to
observation of new optical effects such as the Anderson localization
of light \cite{Wiersma, Storzer}, coherent backscattering in the low
disorder limit \cite{Akkermans} and deviations of statistical
properties of light from the classical diffusion picture
(subdiffusion \cite{Laptyeva} or superdiffusion \cite{Burresi,
Krivolapov}). Traditionally, the realization of optical disordered
system is the strongly scattering medium, such as suspension of
dielectric particles or specially prepared glass. Last years, more
technologically complex systems attract attention, for example, the
``purposely spoiled'' photonic crystals \cite{John, Schwartz,
Reyes-Gomes}, microstructured waveguides \cite{Patterson}, and
metamaterials \cite{Gredeskul}. The study of such artificial
disordered structures is not only intriguing in itself, but also has
great practical value, since the products of real-life
nanofabrication are never perfect.

Introduction of nonlinearity to disordered systems strongly enhances
the complexity and richness of possible optical dynamics. Though
there are many works devoted to this topic, one cannot say that the
problem of disorder/nonlinearity interaction is fully understood.
The results obtained to date include suppression of the Anderson
localization in photonic lattices \cite{Laptyeva, Jovic},
bistability of light transmission through nonlinear random media due
to resonant localized modes \cite{Shadrivov}, instability of waves
in random media with instantaneous and noninstantaneous Kerr
nonlinearity \cite{Skipetrov2000, Skipetrov2003}, influence of
nonlinearity on random laser output \cite{Liu}, etc. Since in this
paper we deal with pulsed radiation propagation, further some
results on this particular topic are briefly discussed.

One of the main features of pulse transmission which allows to make
conclusions on light diffusion and localization is the ``tail'' of
the pulse, i.e. the shape of intensity decrease at the output of the
medium. Exponential tail is the indication of classical diffusion,
while nonexponentiality may evidence energy transfer to long-living
states and, hence, Anderson localization. Appearance of the
nonexponential tail induced by nonlinearity in a three-dimensional
random medium was reported in Ref. \cite{Conti2007}. A number of
works was dedicated to the influence of the so-called transverse
disorder on wave-packet dynamics in two-dimensional nonlinear
structures. In particular, it was shown \cite{Pikovsky} that weak
nonlinearity in a one-dimensional disordered lattice can destroy the
Anderson localization, so that the initially localized wave packet
spreads slower than in the diffusion regime (subdiffusion). On the
other hand, Lahini \textit{et al.} \cite{Lahini} demonstrated that
the packet localization in a similar model described by the discrete
nonlinear Shr\"{o}dinger equation can be stronger or weaker for
different types of modes. This conclusion was confirmed
experimentally using the disordered set of coupled optical
waveguides. On the basis of numerical simulations of wave-packet
dynamics in the one-dimensional lattice, the authors of Ref.
\cite{Flach} distinguished three regimes of evolution: (i)
weak-nonlinearity regime when the packet stays localized for some
time and then spreads subdiffusively, (ii) moderate-nonlinearity
regime when the packet experiences subdiffusion from the very
outset, (iii) strong-nonlinearity regime when some part of the
packet gets localized due to self-trapping. Leaving aside purely
nonlinear process of self-trapping, one can say that any level of
nonlinearity is enough to destroy the Anderson localization. The
effect of self-trapping in disordered systems was studied in later
works as well (see, for example, \cite{Kopidakis, Radosavljevic}).

\begin{figure}[t!]
\includegraphics[scale=0.9, clip=]{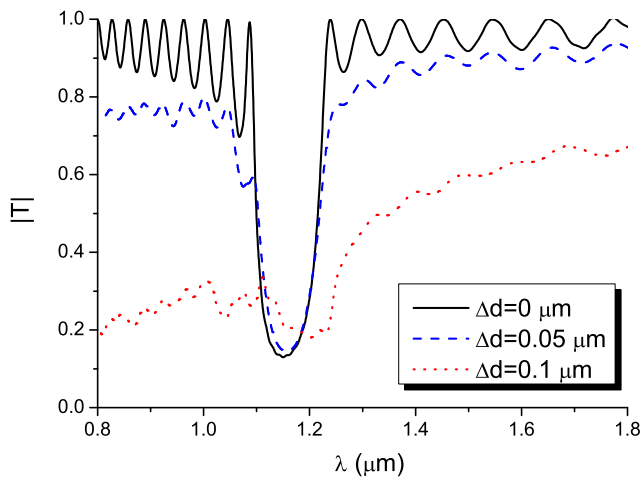}
\caption{\label{fig1} (Color online) Transmission spectra of the
photonic crystal with $10$ periods for different values of disorder
amplitude. The curves are averaged over $20$ realizations.}
\end{figure}

Another consequence of interaction between disorder and nonlinearity
revealed in Ref. \cite{Conti2012} implies formation at certain
conditions of the stable states combining the properties of
localized modes and solitary waves. It was shown that disorder
assists nonlinearity in threshold lowering for generation of such
``superlocalized'' solitons. It was proposed \cite{Folli2011} that
pulsed radiation in the form of self-induced transparency solitons
can be used to pump localized Anderson states in a disordered
two-level medium. Since these Anderson states are similar to the
modes of a laser cavity, the scheme described is expected to operate
as a peculiar two-level laser. Finally, if the nonlinearity is
nonlocal, destabilization of localized states becomes more
problematic: the stronger nonlocality, the higher power needed to
obtain mode instability \cite{Folli2012}.

In this paper, we consider ultrashort pulse propagation in a
one-dimensional disordered multilayer (photonic crystal) with
instantaneous and noninstantaneous nonlinearity. In other words, we
deal not with \textit{transverse}, but with \textit{longitudinal}
disorder. Theoretical study of perfect multilayer systems and their
applications continues for many years (see, for example, Refs.
\cite{Yeh1, Yeh2}). It is known that there is no threshold for the
Anderson localization in one-dimensional random media. The
peculiarity of photonic crystals is their periodicity which results
in different regimes of localization \cite{Vlasov}. We are
interested in the regime of both \textit{strong disorder}, when the
range of Bragg-like reflection is effectively extended beyond the
band gap, and \textit{strong nonlinearity}, when the effects of
pulse shape transformation occur. In Section \ref{instant}, the case
of instantaneous nonlinearity is considered, while the results
taking into account the finite relaxation times of nonlinearity
(noninstantaneous nonlinearity) are discussed in Section
\ref{noninstant}. Separate Section \ref{trap} is devoted to the
self-trapping effect in the relaxing-nonlinearity case. The paper is
closed with a brief conclusion.

\section{\label{instant}Instantaneous nonlinearity}

\begin{figure}[t!]
\includegraphics[scale=0.9, clip=]{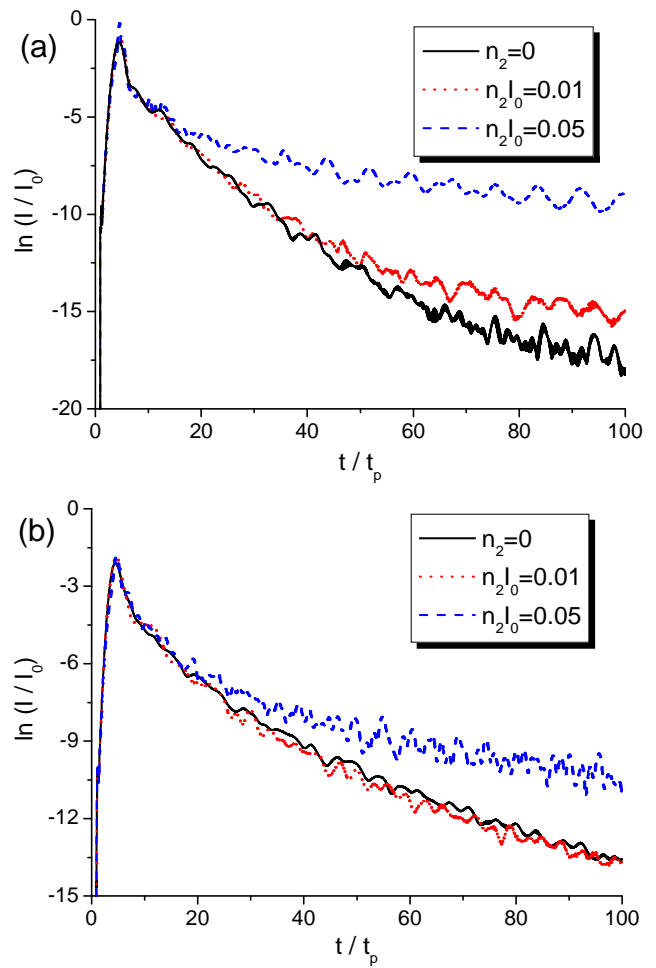}
\caption{\label{fig2} (Color online) Dynamics of transmitted
intensity logarithm for the photonic crystal with $50$ periods at
the disorder amplitudes (a) $\Delta d=0.05$ $\mu$m, (b) $\Delta
d=0.1$ $\mu$m. The curves are averaged over $25$ realizations.}
\end{figure}

\begin{figure}[t!]
\includegraphics[scale=0.9, clip=]{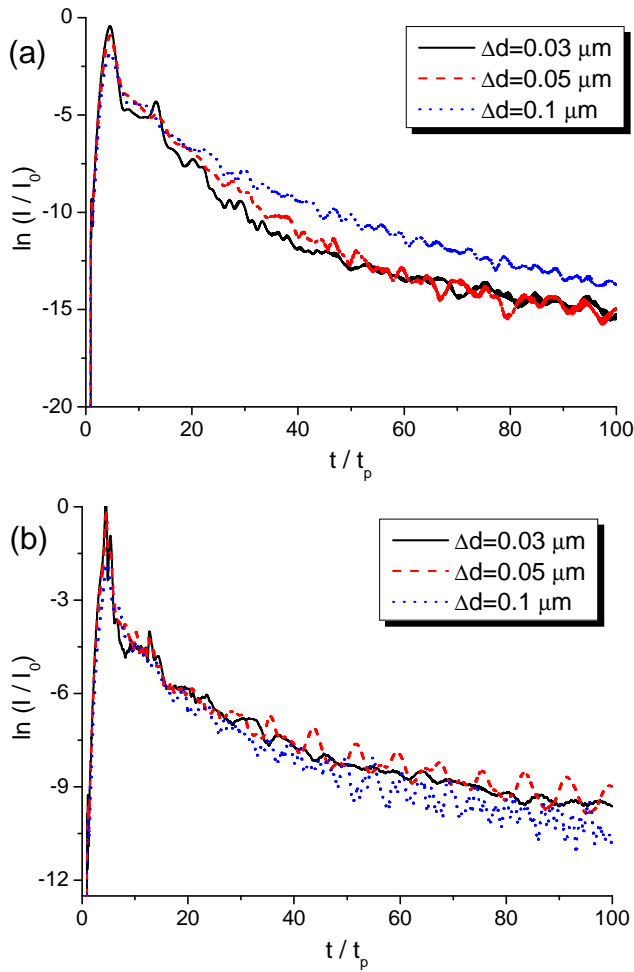}
\caption{\label{fig3} (Color online) Dynamics of transmitted
intensity logarithm for the photonic crystal with $50$ periods at
the nonlinearity coefficients (a) $n_2 I_0=0.01$, (b) $n_2
I_0=0.05$. The curves are averaged over $25$ realizations.}
\end{figure}

\begin{figure}[t!]
\includegraphics[scale=0.85, clip=]{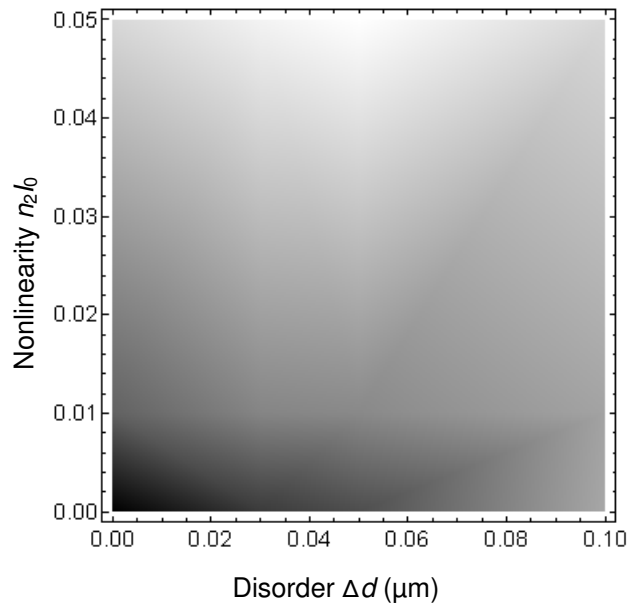}
\caption{\label{fig4} The tail intensity deviation from the linear
ordered case as a function of disorder and nonlinearity strengths.
This density plot is based on the data of Figs. \ref{fig2} and
\ref{fig3}.}
\end{figure}

\begin{figure}[t!]
\includegraphics[scale=0.9, clip=]{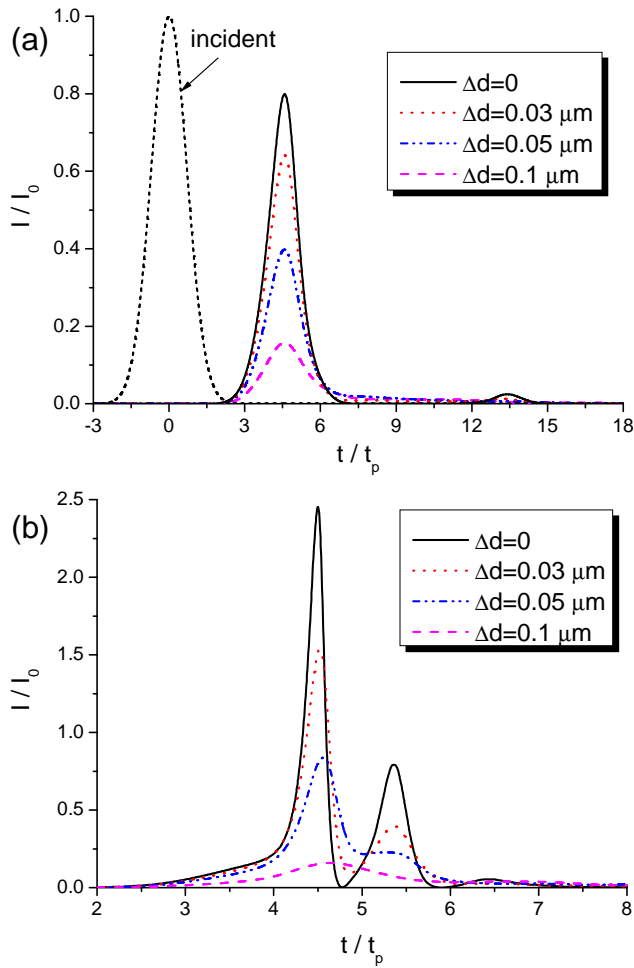}
\caption{\label{fig5} (Color online) The profiles of transmitted
pulses for the photonic crystal with $50$ periods at the
nonlinearity coefficients (a) $n_2 I_0=0.01$, (b) $n_2 I_0=0.05$.
The curves are averaged over $25$ realizations.}
\end{figure}

\begin{figure*}[t!]
\includegraphics[scale=1, clip=]{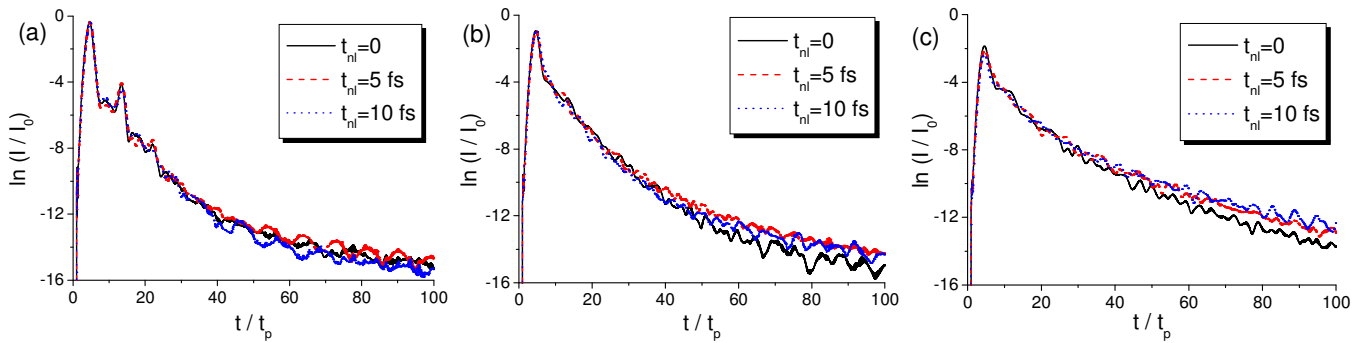}
\caption{\label{fig6} (Color online) Dynamics of transmitted
intensity logarithm for the photonic crystal with $50$ periods at
different relaxation times. The disorder amplitudes are (a) $\Delta
d=0.02$ $\mu$m,(b) $\Delta d=0.05$ $\mu$m, (c) $\Delta d=0.1$
$\mu$m. The nonlinearity coefficient is $n_2 I_0=0.01$. The curves
are averaged over $25$ realizations.}
\end{figure*}

\begin{figure*}[t!]
\includegraphics[scale=1, clip=]{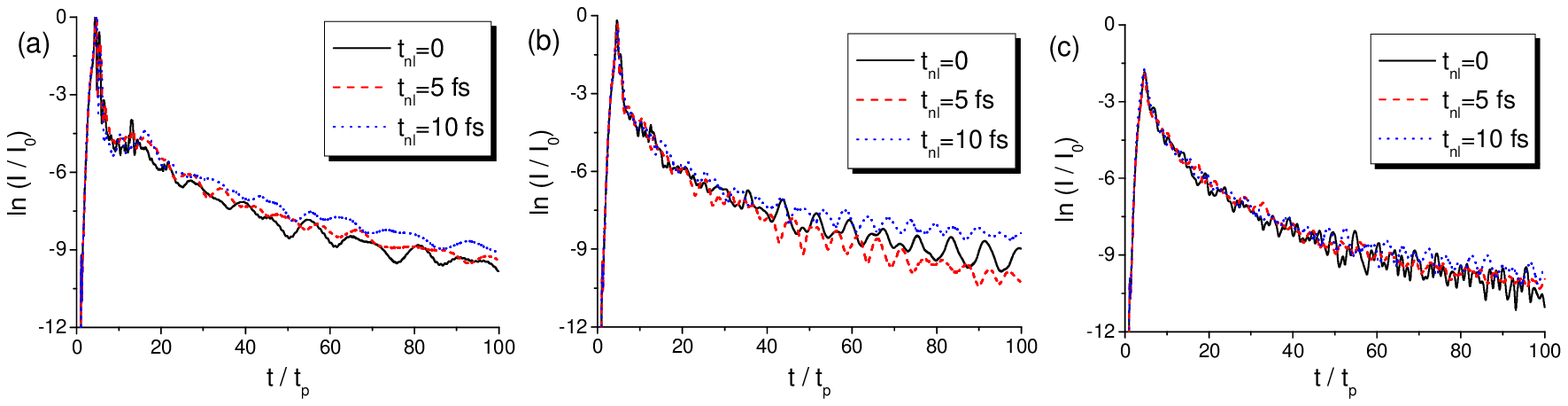}
\caption{\label{fig7} (Color online) The same as in Fig. \ref{fig6}
but for the nonlinearity coefficient is $n_2 I_0=0.05$.}
\end{figure*}

The basis of our calculations is the one-dimensional wave equation,
\begin{eqnarray}
\frac{\partial^2 E}{\partial z^2}&-&\frac{1}{c^2} \frac{\partial^2
(n^2 E)}{\partial t^2} = 0, \label{Max}
\end{eqnarray}
where $E$ is the electric field strength, $n$ is the medium
refractive index. In nonlinear medium, the latter depends on light
intensity $I=|E|^2$ as
\begin{eqnarray}
n=n^0(z)+\delta n (I, t, z), \label{refr}
\end{eqnarray}
where $n^0(z)$ is a linear part of refractive index. The
$z$-dependence marks the periodic modulation inherent for the
photonic crystal which, in our case, is a multilayer structure
consisting of two different materials -- layers $a$ and $b$. In
general, the nonlinear term $\delta n$ takes into account the
relaxation process which we describe with the Debye model of
nonlinearity \cite{Akhm}
\begin{eqnarray}
t_{nl} \frac{d \delta n}{d t}+ \delta n=n_2 I, \label{relax}
\end{eqnarray}
where $n_2$ is the Kerr nonlinear coefficient, and $t_{nl}$ is the
relaxation time. The disorder is introduced into the periodic
structure through random variations of thicknesses of layers $a$ and
$b$,
\begin{eqnarray}
d_{a,b}=d_{a,b}^0+\Delta d (\xi-1/2), \label{rand}
\end{eqnarray}
where $d_{a,b}^0$ are the mean values of thicknesses, $\Delta d$ is
the amplitude of disorder, and $\xi$ is the random quantity
uniformly distributed in the range $[0, 1]$. To solve the equations
listed above, we use the numerical approach developed in one of our
previous publications \cite{Novitsky}.

For our model calculations, we adopt the parameters of the materials
and structure as follows: $d_a^0=0.4$ and $d_b^0=0.24$ $\mu$m,
$n^0_a=2$ and $n^0_b=1.5$. It is worth to note that we do not mean
any specific materials. The main feature of the system considered is
the band gap in optical frequency region which requires layer
thicknesses to be comparable to the wavelength and large enough
difference between refractive indices. The interaction of radiation
with the structure depends on the position of light frequency
against the zone spectrum of the photonic crystal. The change in
these parameters, obviously, will change the position and widths of
the band gaps, so that one would have to make just the proper
frequency tuning. We suppose that it will not lead to any
qualitative changes in the light-matter interaction dynamics. In
this section, we assume that the nonlinearity is instantaneous, i.e.
$t_{nl}=0$. The envelope of the pulse at the input of the photonic
structure has Gaussian shape, $A(t)=A_0 \exp(-t^2/2t_p^2)$, where
$t_p$ is the pulse duration, and $A_0$ is the amplitude of the
radiation field. Let us illustrate influence of disorder on the
properties of photonic crystal with Fig. \ref{fig1} where the
transmission spectra of the structure with $N=10$ periods are
depicted. It is seen that strong disorder results in decrease of
transmission outside the initial band gap. This can be interpreted
as effective widening of the band gap in accordance with Ref.
\cite{Vlasov}. The waves attenuate exponentially while propagating
in this widened forbidden region which is the feature of
localization effect. For definiteness, we assume $t_p=50$ fs and the
central wavelength $\lambda_c=1.064$ $\mu$m in further calculations,
i.e. the carrier frequency lies just outside the band gap of the
perfect multilayer.

In this paper, we restrict ourselves to the structure with nonlinear
$b$ layers (second layers of the period), since light concentrates
in these layers when we deal with the high-frequency edge of the
band gap. There is a number of studies devoted to such semi-linear
structures which can be used, for example, for optical nonlinearity
management \cite{Centurion}. The strength of nonlinearity ($n_2
I_0=n_2 |A_0|^2 \sim 0.01$) is taken to be large enough to strongly
influence the pulse characteristics. In addition, it allows to
consider comparatively short systems, in this section we assume
$N=50$ periods. Thus, we are interested in interplay of strong
disorder and strong nonlinearity. This interplay can be studied on
the short timescale (pulse shape transformation) and at long times
(pulse tail transformation).

Let us start with the change of pulse decay pattern in the nonlinear
structure. As was mentioned above, nonexponential behavior of the
tail can be interpreted as the evidence of diffusion violation and
the sign of the Anderson localization. In other words, the logarithm
of intensity should be a linear function of time in the diffusion
regime. As seen in Fig. \ref{fig2}, this dependence deviates from
linear even in the linear case ($n_2=0$), since there is no
threshold for localization. Nonlinearity leads to stronger
nonexponentiality which can be treated as light induced localization
\cite{Conti2007}. In other words, the number of photons traveling in
the structure for a long time before eventually leaving it
(long-lived localized states) is larger in the nonlinear case than
in the linear one. Comparing Figs. \ref{fig2}(a) and (b), one can
say that the larger disorder, the stronger nonlinearity is needed to
observe the deviation from the linear tail. Figure \ref{fig3} shows
that the opposite is also true: the large nonlinearity coefficient
implies the necessity to make disorder stronger than at lower
nonlinearities to obtain the deviation from the nonlinear ordered
system behavior. It is seen that in the case of $n_2 I_0=0.05$, the
rise of disorder almost has no influence on the rate of intensity
decay. Thus, nonlinearity and disorder seem to be the
\textit{competing} factors: strong disorder suppresses the
manifestations of nonlinearity at large times, and vice versa. On
the other hand, both factors act in the \textit{same} direction:
they tend to increase the nonexponentiality of the tail.

The results of Figs. \ref{fig2} and \ref{fig3} are summed up in Fig.
\ref{fig4}: the change of grey level from black to white shows the
deviation of the tail intensity (at $t=100 t_p$) from the linear
ordered case. It is seen that there is optimal value of disorder
($\Delta d=0.05$ $\mu$m) at which the influence of the strong
nonlinearity on the pulse dynamics is maximal. When both these
factors are very strong (right upper corner of the diagram), the
resulting effect is not so pronounced. It is worth to emphasize that
here we compared the results with the single linear ordered case,
while in Figs. \ref{fig2} and \ref{fig3} we considered the
deviations from the behavior of the corresponding linear disordered
and nonlinear ordered systems.

Let us now proceed from the long timescale where the effects like
localization can appear to short times and study the change of pulse
form under joint influence of nonlinearity and disorder. The results
of calculations for the two values of nonlinearity are demonstrated
in Fig. \ref{fig5}. In the case of smaller value of nonlinearity
[Fig. \ref{fig5}(a)], disorder results only in transmission
attenuation and intensification of reflection. The peak of the pulse
propagates with the same speed for all values of $\Delta d$. Strong
nonlinearity is the reason of such effects as compression and
dissociation of the pulse [Fig. \ref{fig5}(b)]. Disorder smoothes
out and suppresses these effects, though at the smaller values of
$\Delta d$ the transmitted pulse is still shorter than the initial
one. As the disorder parameter grows, the shape of the pulse tends
to be more and more symmetric. Moreover, the peak of the transmitted
pulse shifts to longer times. In other words, strong nonlinearity
and disorder act \textit{together} to slow down the pulse. Thus,
both on the long and short timescale, the disorder and nonlinearity
are the factors acting in the same direction in one respect and in
the opposite directions in another respect.

\section{\label{noninstant}Relaxing nonlinearity}

\begin{figure}[t!]
\includegraphics[scale=0.9, clip=]{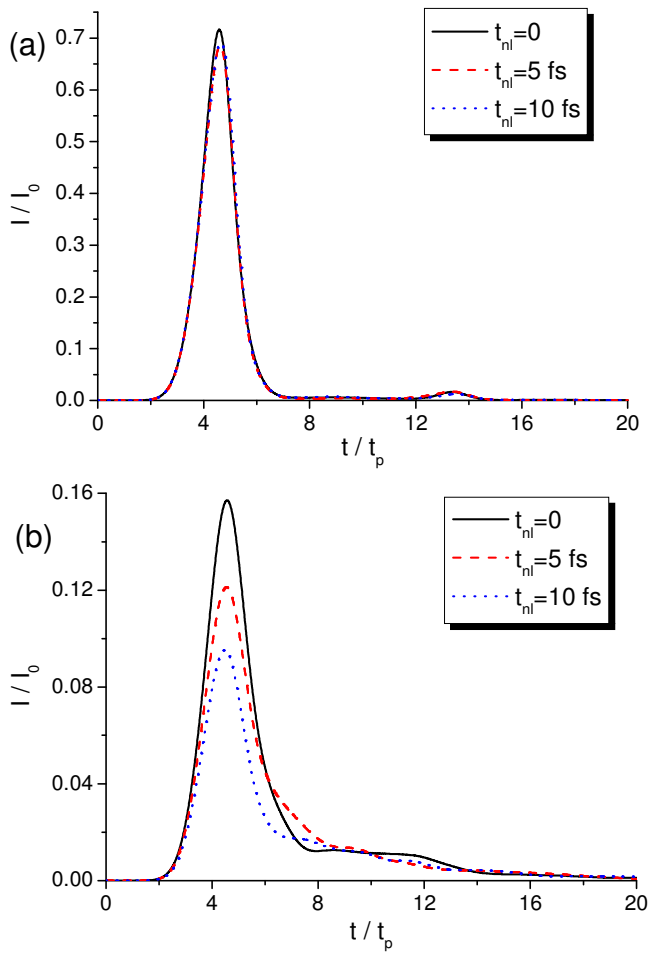}
\caption{\label{fig8} (Color online) Profiles of the pulses
transmitted through the photonic crystal with $50$ periods at
different relaxation times. The disorder amplitudes are (a) $\Delta
d=0.02$ $\mu$m, (b) $\Delta d=0.1$ $\mu$m. The nonlinearity
coefficient is $n_2 I_0=0.01$. The curves are averaged over $25$
realizations.}
\end{figure}

\begin{figure}[t!]
\includegraphics[scale=0.9, clip=]{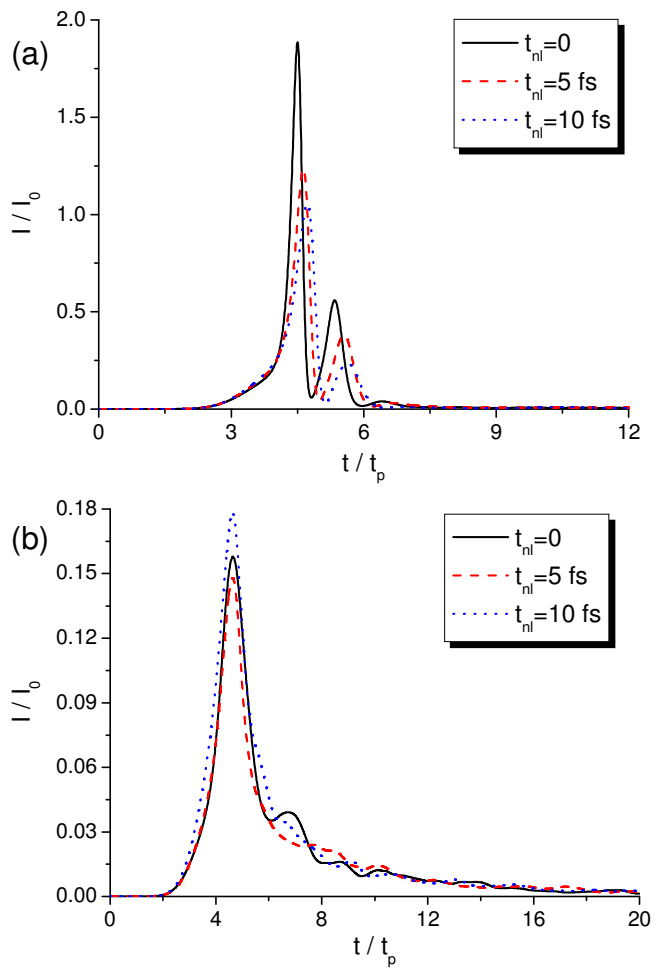}
\caption{\label{fig9} (Color online) The same as in Fig. \ref{fig8}
but for the nonlinearity coefficient is $n_2 I_0=0.05$.}
\end{figure}

Now let us turn to the noninstantaneous nonlinearity assuming that
the relaxation time in Eq. (\ref{relax}) is not zero. In our
calculations, the values $t_{nl}=5$ and $10$ fs are used which
approximately correspond to the electronic mechanism of relaxation;
the pulse duration is still $50$ fs. First, we discuss the results
for the long timescale, i.e. the change of behavior of the tail due
to the appearance of the relaxation. The case of relatively weak
nonlinearity is shown in Fig. \ref{fig6}. Panel (a) of this figure
demonstrates that, for both weak disorder and weak nonlinearity, the
temporal dependence of the tail is unaffected by the relaxation
process. Increasing disorder strength results in divergence of the
curves: nonexponentiality of the tail increases for nonzero $t_{nl}$
[see Fig. \ref{fig6}(b) and (c)]. This can be considered as
indication of stronger Anderson localization for relaxing weak
nonlinearity at large disorder parameters.

The results for strong nonlinearity are depicted in Fig. \ref{fig7}.
This figure implies that the increase of disorder has the opposite
effect: the curves for different relaxation times take on similar
behavior for large $\Delta d$ [Fig. \ref{fig7}(c)], while for weak
disorder they follow substantially different paths [Fig.
\ref{fig7}(a) and (b)]. Thus, in the cases when both disorder and
nonlinearity are strong or weak, the influence of relaxation is
negligible on the long timescale. To see the change of behavior due
to $t_{nl} \neq 0$, one should take either strong disorder and weak
nonlinearity or vice versa.

On the short timescale, we are interested in the change of the peak
intensity of the pulse under simultaneous presence of
noninstantaneous nonlinearity and disorder. As pulse profiles in
Figs. \ref{fig8} and \ref{fig9} show, the relationship between the
factors is the same as on the long timescale. The influence of
relaxation is minimal when both nonlinearity and disorder are weak
[Fig. \ref{fig8}(a)]. The effect of relaxation is maximal when one
of the factors is strong while the other is weak [Fig. \ref{fig8}(b)
and \ref{fig9}(a)]. Finally, when both nonlinearity and disorder are
strong, the change in pulse profiles is not great again [Fig.
\ref{fig9}(b)], though in this case increase of relaxation time can
even result in the rise of peak intensity (see the curve for
$t_{nl}=10$ fs).

\section{\label{trap}Self-trapping}

\begin{figure}[t!]
\includegraphics[scale=0.9, clip=]{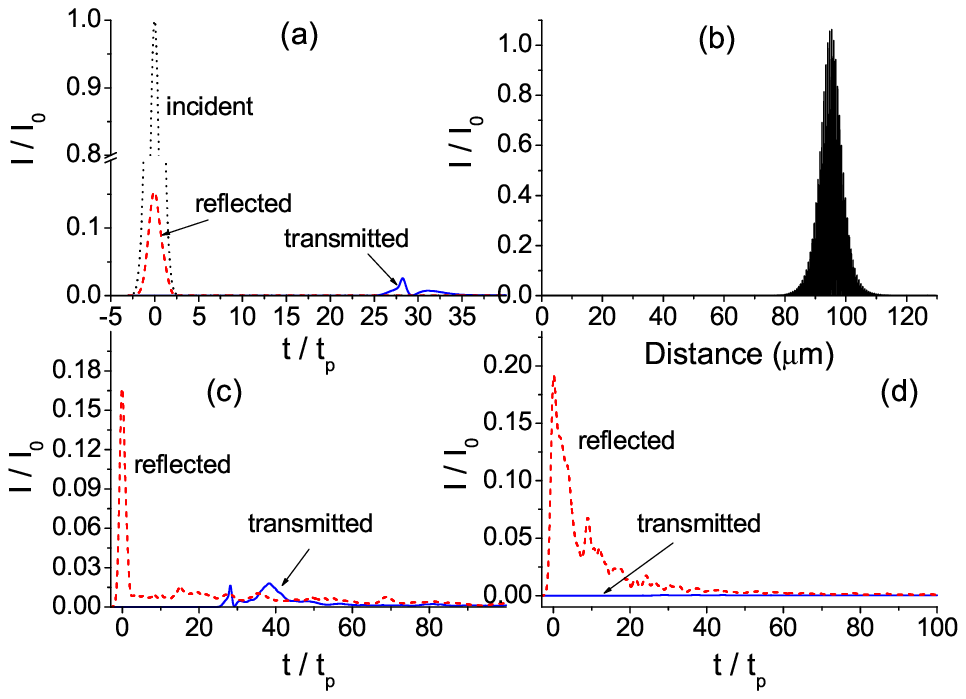}
\caption{\label{fig10} (Color online) (a), (c), (d) Profiles of the
transmitted and reflected radiation when the disorder amplitudes are
$\Delta d=0$, $0.02$ and $0.1$ $\mu$m, respectively. (b) The
distribution of light intensity inside the structure at the time
instant $t=100 t_p$ in the ordered case. The photonic crystal has
$200$ periods (length $~128$ $\mu$m); the nonlinearity coefficient
is $n_2 I_0=0.01$ and the relaxation time is $t_{nl}=10$ fs. The
curves in panels (c) and (d) are averaged over $25$ realizations.}
\end{figure}

\begin{figure}[t!]
\includegraphics[scale=0.9, clip=]{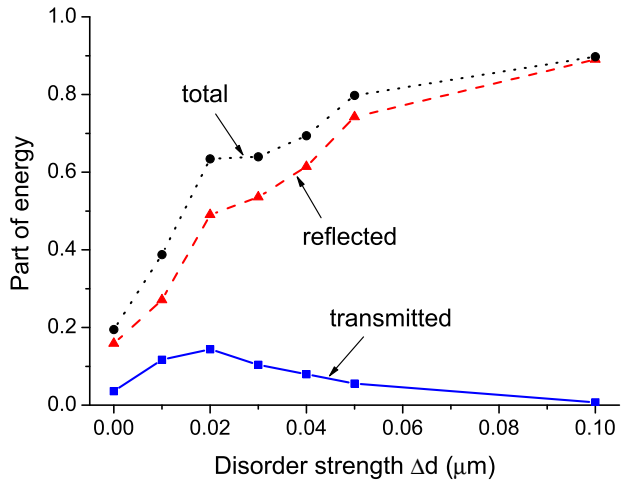}
\caption{\label{fig11} (Color online) The dependence of the output
energy (as a part of the input one) on the disorder strength. The
curves are the result of averaging over $25$ realizations.}
\end{figure}

\begin{figure}[t!]
\includegraphics[scale=0.9, clip=]{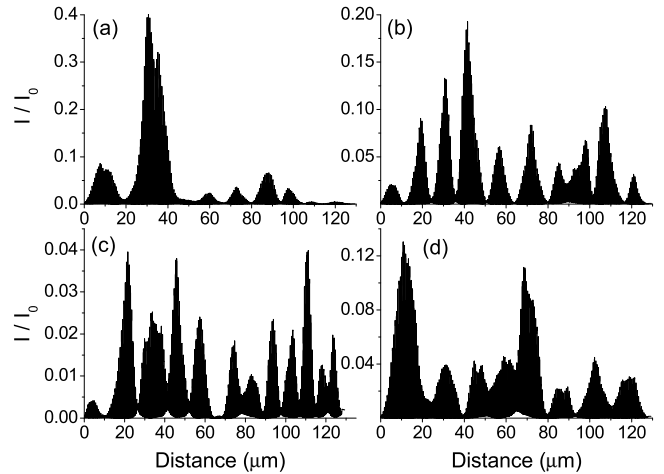}
\caption{\label{fig12} Four realizations of the distribution of
light intensity inside the structure at the time instant $t=100 t_p$
in the case $\Delta d=0.02$ $\mu$m.}
\end{figure}

\begin{figure}[t!]
\includegraphics[scale=0.9, clip=]{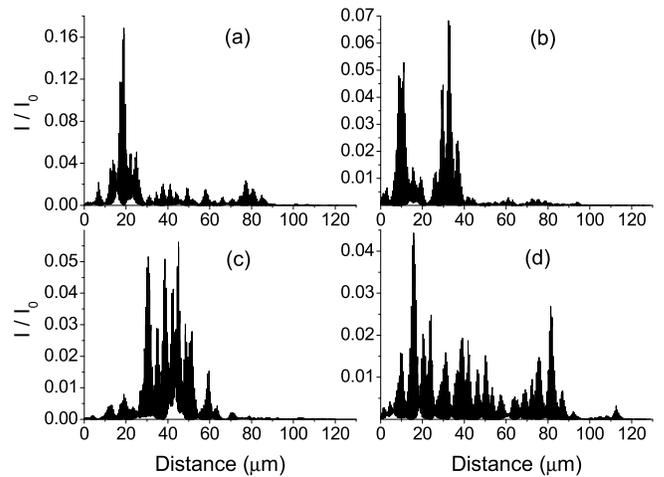}
\caption{\label{fig13} The same as in Fig. \ref{fig12} but for the
disorder strength $\Delta d=0.1$ $\mu$m.}
\end{figure}

In this section, we study the influence of disorder on the
self-trapping of the pulse in the nonlinear photonic crystal.
Self-trapping occurs when both the pulse duration and the relaxation
time of nonlinearity are short enough and comparable with each
other. The reason is the formation of a peculiar dynamical cavity
(distribution of nonlinear index of refraction) which is stable
enough to trap the energy of the pulse for a long time. Different
aspects of the self-trapping effect were studied in our previous
papers \cite{Novitsky, Novitsky1, Novitsky2, Novitsky3, Novitsky4}.
To obtain self-trapping, in this section we assume shorter pulse
($t_p=30$ fs) and longer photonic crystal ($N=200$ periods). As
previously, the half of the layers is linear, while the other half
has relaxing nonlinearity. First, we assume the disorder is absent.
The profiles of the transmitted and reflected radiation
demonstrating self-trapping at $t_{nl}=10$ fs are shown in Fig.
\ref{fig10}(a). One can easily see that the energy of the output
radiation in this case is much smaller that the input energy, i.e.
the most part of light is stored inside the structure. The
distribution of radiation intensity over the system at the time
instant $t=100 t_p$ (more than three times larger than the
propagation time of the pulse in the linear case) has the
characteristic bell-shaped form resembling the envelope of the pulse
[see Fig. \ref{fig10}(b)].

Now we can add the disorder. The dependence of the output energy
integrated for the time $100 t_p$ on the disorder strength is shown
in Fig. \ref{fig11}. It is seen that the dependence can be divided
into three regions:

(i) the region of rapid rising where both the transmitted and
reflected energies increase. This region occupies the disorder
strengths from $\Delta d=0$ [order, when the transmission is low and
the reflection is minimal, see Fig. \ref{fig10}(a)] to $\Delta
d=0.02$ $\mu$m [transmission has maximal value, see the profiles in
Fig. \ref{fig10}(c)];

(ii) the region where the transmission curve rapidly decreases while
the reflection curve has a depression. This region is between
$\Delta d=0.02$ and $0.05$ $\mu$m;

(iii) the region of gently sloping curves for $\Delta d>0.05$ $\mu$m
where transmission is negligible and reflection is very strong [see
the profiles in Fig. \ref{fig10}(d)].

Thus, the main effect of disorder is the gradual increase of
reflection, while the transmission has the maximum at a certain
level of disorder. This behavior can be explained as follows. First,
at low disorder strengths ($\Delta d<0.02$ $\mu$m), the dynamical
trap (or cavity) loses its stability and more radiation leaves the
structure during the same time. As a result, both the transmission
and reflection grow. However, more than a half of light energy stays
inside the system. Figure \ref{fig12} shows four typical
realizations of light distribution inside the disordered photonic
crystal at the point of maximal transmission ($\Delta d=0.02$
$\mu$m). Rare realizations which appear once in the full set of
realizations have a single high-intensity peak corresponding to the
trap [Fig. \ref{fig12}(a)] or many low-intensity peaks distributed
over the length of the structure [Fig. \ref{fig12}(c)]. Between
these extremes are the cases of many narrow peaks [Fig.
\ref{fig12}(b)] or several wide peaks of intermediate intensity
[Fig. \ref{fig12}(d)]. It is worth to note that the high-intensity
sites of the distribution are concentrated mainly close to the input
of the system, in contrast to the ordered case [compare with Fig.
\ref{fig10}(b)].

For higher disorder strengths ($\Delta d>0.02$ $\mu$m), the process
of reflection reinforcement typical for strongly disordered media
becomes dominant. At the same time, transmission rapidly decays. As
distribution realizations at $\Delta d=0.1$ $\mu$m show (Fig.
\ref{fig13}), the remanent radiation concentrates almost exclusively
in the first half of the photonic crystal. Possible variants include
a single narrow high-intensity peak near the input facet [Fig.
\ref{fig13}(a)], two peaks of intermediate intensity [Fig.
\ref{fig13}(b)], many peaks in a narrow region [Fig.
\ref{fig13}(c)], and, finally, a wide distribution of low-intensity
radiation covering almost two third of the structure [Fig.
\ref{fig13}(d)]. Thus, the maximum in the transmission curve of Fig.
\ref{fig11} is a result of two processes occurring due to disorder:
reflection reinforcement and destruction of the trap.

\section{Conclusion}

In this paper, we analyzed the interplay of disorder and
nonlinearity in several variants of one-dimensional multilayer
structure. Calculations for both instantaneous and relaxing
nonlinearity imply that nonlinearity and disorder are the competing
factors: when one of them is strong, the influence of the other
factor becomes apparent at substantially larger values than in the
weak-first-factor case. In addition, in the regime of strong
disorder and strong nonlinearity, the influence of relaxation
appears to be negligible. Special attention was given to the
modification of the self-trapping effect which still takes place for
low enough disorder strengths. However, the patterns of light
distribution along the photonic crystal are substantially changed as
demonstrated by various realizations of disordered system. At higher
disorders, the usual reflection prevails diminishing the part of
energy trapped inside the structure.

Possible further directions of research include studies of spectral
transformations and multi-pulse dynamics inside disordered photonic
crystals. It is also important to increase the number of
realizations. Though the qualitative conclusions based on the
general trends of the curves seem to be reliable, the number of
realizations needs to be improved in future.

Though the results reported in this paper were obtained for optical
waves, we hope that they will be useful for the other physical
situations as well, such as nonlinear acoustic waves, and are of
interest from the wider perspective of general classical wave
dynamics. Moreover, nonlinearity of the medium can be considered in
some sense as an analogue of electron-electron interaction in the
solid state disordered systems \cite{Skipetrov2003a}.

\acknowledgements{This work was supported by the Belarusian State
Foundation for Basic Research, project F13M-038.}

\end{document}